# TRPC4 ion channel protein is selectively expressed in a subpopulation of dopamine neurons in the ventral tegmental area

Kurt R. Illig[1], Kristin C. Rasmus[2], Andrew L. Varnell[2], Eric M. Ostertag[3], William D. Klipec & Donald C. Cooper[2*]

**The nonselective cation channel TRPC4 has been shown to be present in high abundance in the corticolimbic regions of the brain and play a pivotal role in modulating cellular excitability due to their involvement in intracellular $Ca^{2+}$ regulation. Recently we reported their involvement in socialization and regulating anxiety-like behaviors in rats. Given the important role for dopamine in modulating emotions involved in social anxiety we investigated whether TRPC4 protein and mRNA was found on dopaminergic neurons of the ventral tegmental area (VTA). Using emulsion autoradiography we found that TRPC4 mRNA is indeed present in the VTA and the substantia nigra. Additionally, immunohistochemistry verified it's presence on a subpopulation of dopamine neurons in the VTA. We confirmed these findings by testing Trpc4 knock-out rats in addition to wild-type animals. This novel finding suggests that TRPC4 plays a pivotal role in regulating dopamine release in a subpopulation of neurons that may modulate emotional and cognitive responses in social situations.**

The TRPC family of nonselective channels are made up of seven members (TRPC1- 7) and the TRPC4 channel is one of the two most abundant in the mammalian brain[1]. Our recent findings indicate that the TRPC4 channel is important for sociability in a rodent model of social interaction[3]. For example, Trpc4 knock-out rats exhibit significantly decreased social interaction compared to their wild-type counterparts[3]. Previous experiments in our lab indicate that TRPC4 channels are highly expressed in the corticolimbic regions including the lateral septum, hippocampus, prefrontal cortex (PFC) and the amygdala[3]. These brain regions receive extensive input from dopamine (DA) neurons in the VTA. Given the importance of DA in modulating reward and stress we tested for the presence of TRPC4 protein in the cells of the VTA. We were particularly interested in whether TRPC4 channels were found in GABAergic interneurons and projection neurons or within tyrosine hydroxylase (TH) expressing DA neurons.

## RESULTS

### *Trpc4 Knock-out Rats*

The Sleeping Beauty (SB) gene-trap transposon method was used to create the Trpc4 knock-out animals[4]. The SB method uses cut-and-paste transposable elements to generate heritable loss-of-function mutations. Fig. **1a** shows the location of the *trpc4* gene on the rat genome and where the transposon was inserted. By inserting the SB transposon into the first intron of the *trpc4* gene, the full-length protein product is completely eliminated. Using primers for the Trpc4 knock-out and wildtype alleles, we were able to confirm the deletion using PCR and gel electrophoresis (Fig **1b**).

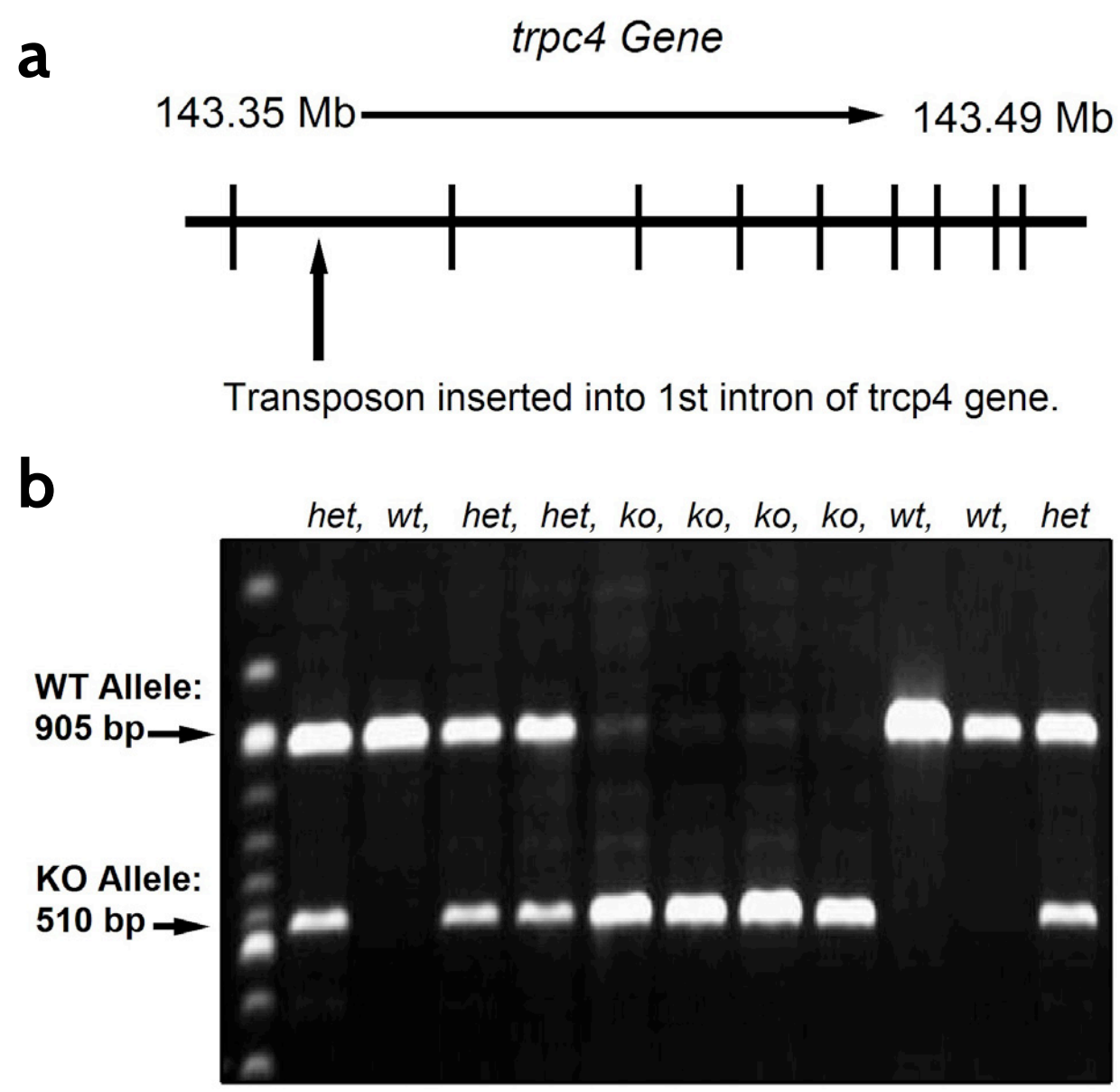

Fig. 1: **a.** Schema of the *trpc4* gene in the rat genome and the Sleeping Beauty (SB) gene knock-out system. The *trpc4* gene is located on chromosome 2 of the rat genome, between 143.35 Mb and 143.49 Mb. The Sleeping Beauty transposon was inserted into the first intron of *trpc4*, therefore creating a complete knock-out of the coding sequence. **b.** Ethidium bromide-stained agarose gel visualizing the 905 bp marker for the WT allele and the 510 bp marker for the *trpc4* KO allele. To genotype the animals, a 1.5% agarose gel electrophoresis was used.

### *TRPC4 Localization*

As shown Fig. 2, TRPC4 expression was found on a subpopulation of dopamine neurons in the VTA, in addition to positive expression in the substantia nigra. TRPC4 could be found on a subpopulation of tyrosine hydroxylase (TH)-expressing cells in the VTA of wild-type rats (Fig. 2G-J). TRPC4 labeling was absent from TH-expressing cells in TRPC4

1.Department of Biology and program in Neuroscience, University of St. Thomas, Saint Paul, Minnesota, 55105
2.Center for Neuroscience, Department of Neuroscience, University of Colorado at Boulder, Boulder Colorado 80303, USA
3.Transposagen Biopharmaceuticals, Inc Lexington, KY 40506
4.Department of Psychology, Drake University, Des Moines, IA 50311, USA
*Correspondence should be sent to D.Cooper@Colorado.edu

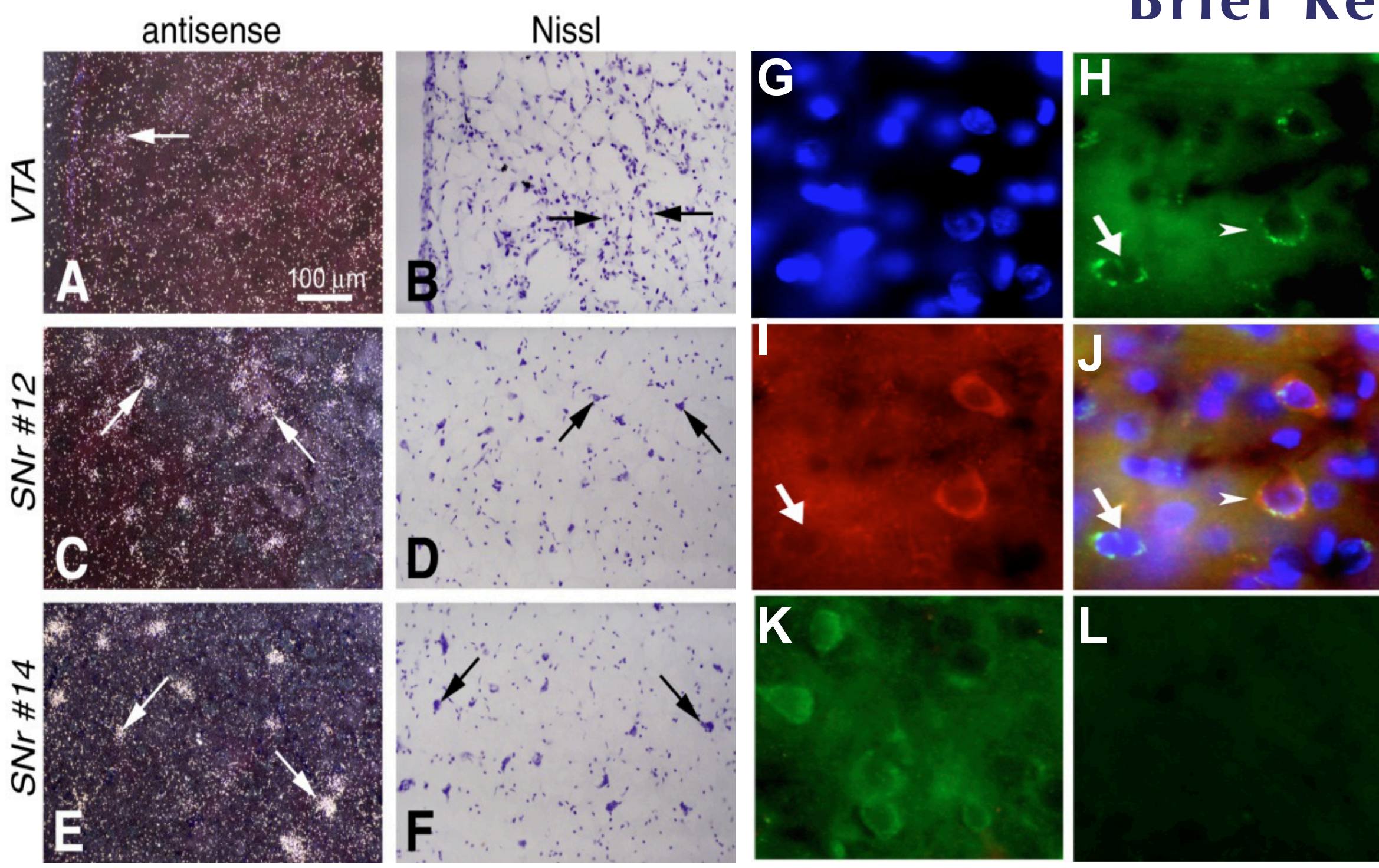

Fig. 2. TRPC4 expression in substantia nigra and VTA. Left Panel: Emulsion autoradiography **A**. Arrow indicates Trpc4 mRNA in rare cells in the ventral tegmental area. **B**. The section near to (**A**) with Nissl staining. Small-size neurons are indicated (arrows). **C**. ISH labeling in substantia nigra reticulata, cell group #12 (arrows). **D**. The section near to (C) with Nissl staining. Group #12 medium-size neurons are indicated (arrows). **E**. ISH labeling in substantia nigra reticulata, cell group #14 (arrows). **F.** The section near to (E) with Nissl staining. Group #14 medium-size neurons are indicated (arrows). Magnification x130. Right Panel: Immunohistochemistry **G-L**: TH and TRPC4 protein expression in WT and KO animals. Tissue from WT animals was triple-labeled for DAPI (**G**), TH (**H**) and TRPC4 (**I**). **J**. Composite of DAPI, TH and TRPC4 labeling in the VTA of a WT rat. **K**. TH and TRPC4 double-labeling in a TRPC4 KO rat. **L**. Control section without primary antibody incubation. Arrows indicate TH-only cells. Arrowheads indicate TH/TRPC4 double-labeled cells**.**

knock-out rats (Fig. 2K). Elimination of the primary antibody in wild-type animals abolished all immunolabeling (Fig. 2L). Fig. **2J** shows a composite of TH and TRPC4 expression in the VTA. The arrow indicates the presence of TH while the arrowhead indicates co-expression of TH and TRPC4.

## DISCUSSION

The localization of TRPC4 in the adult rat brain is consistent with previous in situ mRNA[1]. In this study we increased the resolution of the TRPC4 mRNA and protein to the cellular level and discovered selective co-localization of TRPC4 channels in subpopulations of TH-positive DA neurons. Although we still lack a clear understanding of the TRPC channel signaling cascade in DA neurons it is likely that these channels contribute to increasing DA neuron excitability. The presence of TRPC4 channels in select subpopulations of DA neurons is intriguing and may indicate a circuit specific modulation of DA activity which could influence reward and anxiety/stress-like behaviors.

## METHODS

### *Animal Genotyping and Quantitative PCR*

Three PCR primers were used and designed as 20-24 oligonucleotide sequences. Reactions were carried out using Choice Taq Blue DNA polymerase in a Techne Touchgene thermal cycler. Ethidium bromide-stained agarose gels were photographed with a UV transilluminator imager.

### *Emulsion Autoradiography*

Labeled cRNA antisense and sense probes were freshly prepared and used. In situ hybridization was carried out according to procedures described previously[3]. Slides were dehydrated and apposed to x-ray film for 5 days and then covered with photographic emulsion for 15 days and silver grains developed with D19 Kodak photographic solution.

### *Immunohistochemistry*

Brain sections were double-labeled for TRPC4 and Tyrosine hydroxylase using fluorescence immunohistochemistry using procedures described previously[6]. Primary antibodies were used prior to incubation. Post incubation, fluorescence-tagged secondary antibodies were applied and incubated again. Brain sections were mounted on slides with ProLong anti-fade reagent containing DAPI. Omission of the primary antibody during processing eliminated all tissue staining.

## PROGRESS AND COLLABORATIONS

To see up to date progress on this project or if you are interested in contributing to this project visit: http://www.Neuro-Cloud.net/nature-precedings/illig/

## ACKNOWLEDGEMENTS

This work was supported by National Institute on Drug Abuse grant R01-DA24040 (to D.C.C.), NIDA K award K-01DA017750 (to D.C.C.).

## AUTHOR CONTRIBUTIONS

K.R.I. and D.C.C. designed the experiments. K.R.I carried out the experiments. K.C.R., A.L.V. and D.C.C. wrote and prepared the manuscript.

1. Fowler MA, et al., *PLoS ONE.* **2(6): e573** (2007)
2. Schaefer M, et al. *J. Bio. Chem.* **275,** 17517-17526, (2000)
3. Rasmus K, et al, *Nat. Pre.* <http://dx.doi.org/10.1038/npre.2011.6367.1> (2011),
4. Geurts AM, et al. *BMC Biotechnol.* **6(30)**. (2006)
5. Paxinos, G. *Academic Press.* **pg 355.** (1985)
6. Illig KR, et al., *J Comp Neurol.* **457(4)***:* 361-73. (2003)